\documentclass[twocolumn]{revtex4-2}
\usepackage{amsmath}
\usepackage{graphicx,xcolor}
\usepackage{bbold}
\usepackage{amsmath,amssymb,amsfonts}
\usepackage{hyperref}
\usepackage{siunitx}
\hypersetup{colorlinks=true,allcolors=blue}

\newcommand{\TUdo}{Condensed Matter Theory, Department of Physics, TU Dortmund, 44221 Dortmund, Germany}

\begin{document}
	\title{Factorization rule for multitime correlations in non-Markovian
		open quantum systems}
	\author{Thomas K. Bracht}
	\email{thomas.bracht@tu-dortmund.de}
	\affiliation{\TUdo}    
	\author{Moritz Cygorek}
	\email{moritz.cygorek@tu-dortmund.de}
	\affiliation{\TUdo}
	
	\begin{abstract}
		Experiments performed on quantum systems often measure multitime correlation functions. When quantum systems are weakly coupled to their environment, the time evolution of such correlation functions can be reduced to that of the reduced density matrix by the quantum regression theorem (QRT). While no QRT is available for general non-Markovian open quantum systems, we show that for time-independent Hamiltonians and finite memory times $\tau_c$, an exact factorization rule exists that relates higher-order multitime correlations to products of lower-order correlations. Consequently, all information needed to reconstruct $n$-time correlations is contained in a temporal volume of $\mathcal{O}(\tau_c^n)$. On the example of quantum dots coupled to phonons, we demonstrate that this factorization makes numerical calculations of multitime correlations extremely efficient and even enables semianalytical solutions  in systems where the standard QRT breaks down.
	\end{abstract}
	\maketitle
	
	While the state of a quantum system is completely defined by its density matrix, experiments often measure response functions, which are described by multitime correlation functions~\cite{glauber1963quantum,Eberly,mcalister1997ultra}. For open quantum systems weakly coupled to environments, the quantum regression theorem (QRT)~\cite{QRT_Lax1963,QRT_Swain,budini2008operator,khan2022quantum} allows one to use the same dynamical map that describes the time evolution of the density matrix also for the evolution of multitime correlations, which tremendously simplifies analysis and simulation. However, the conditions for the validity of the QRT are rather restricted and does not even fully cover the class of Markovian open quantum systems~\cite{QRT_Guarnieri} as defined by contractivity of the trace distance~\cite{BLP} or by the divisibility of the dynamical map~\cite{RHP, chruscinski2022dynamical}. It relies instead on the Born approximation, i.e., the assumption that system and environment remain uncorrelated~\cite{QRT_Swain,stauber2000electronphonon}.
	
	This condition is not fulfilled in technologically important classes of quantum devices, such as bright and pure non-classical light sources based on semiconductor quantum dots (QDs)~\cite{akopian2006entangled,lodahl2016interfacing,heindel2023quantumdots,twophoton,huber2018semiconductor}, where strong coupling to longitudinal acoustic phonons~\cite{krummheuer2005pure,reiter2019distinctive,reiter2014roleofphonons} is known to result in non-Markovian dynamics \cite{rossi2002theory,ilessmith2017limits,ilessmith2017phonon} including the breakdown of the conventional QRT~\cite{nonMarkovian_McCutcheon,QRT_Cosacchi}. This is especially deplorable for quantum emitters as most figures of merit of the emitted light are defined by multitime correlations~\cite{Kiraz,mcalister1997ultrafast,bracht2024theory,delvalle2012theory}. These may be calculated, e.g., by brute-force numerically exact path integral techniques~\cite{PImultitime}, or, for certain correlation functions, by reformulating them as operator averages~\cite{alonso2005multiple,IlesSmith2026,bundgaardnielsen2026quantum}.
	
	The lack of a QRT for strongly coupled open quantum systems~\cite{QRT_FordOConnell} naturally raises the questions whether  the time evolution of multitime correlations contains genuinely different information than the dynamical maps describing the evolution of the density matrix, or whether there exist relations between the two. Such relations could facilitate the efficient evaluation of multitime correlations of non-Markovian open quantum systems and thus circumvent the need for a QRT for general open quantum systems.
	
	Here, we show that there is an exact factorization rule that allows breaking down $n$-th order correlations into lower-order correlations provided (i) the Hamiltonian is independent of time, (ii) the memory time $\tau_c$ (as defined in Ref.~\cite{timelocal}) is finite, and (iii) subsequent time arguments are separated by at least the memory time $t_{i+1}-t_i\ge\tau_c$. This implies that the complete information about the largest unfactorizable $n$-time correlation is contained in a temporal volume of $\tau_c^n$.
	
	Our derivation of the factorization rule (see Appendix~\ref{app:deriv}) is based on describing the dynamics in the eigenbasis of the time-independent Liouvillian of a Markovian embedding. Many eigenstate contributions within this embedding quickly decay until---by definition of the memory time $\tau_c$~\cite{timelocal}---only about as many remain as there are degrees of freedom in the reduced system. The resulting one-to-one correspondence between reduced system degrees of freedom and the surviving Liouvillian eigenstate contributions in the Markovian embedding make it possible to encode the exact dynamics within the reduced system degrees of freedom alone. As a consequence, the dynamical map $\rho(t)=\mathcal{E}_{t,t_0}\rho(t_0)$ expressed in its time-local form  $\mathcal{E}_{t+\Delta t, t}=\mathcal{E}_{t+\Delta t, t_0} \mathcal{E}_{t,t_0}^{-1}$ for time step $\Delta t$ becomes stationary (independent of time $t$) after times $t>t_0+\tau_c$ exceeding the memory time of the environment~\cite{timelocal_Clark,timelocal}. 
	
\begin{figure}
		\begin{center}
			\includegraphics[width=\linewidth]{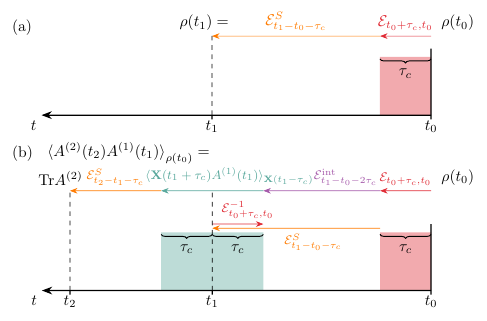}
		\end{center}
		\caption{\label{fig:fac_two}(a) The dynamical map for density matrix evolution becomes stationary after the memory time $t\ge\tau_c$, which enables exact long-time extrapolation of non-Markovian dynamics. (b) Exact factorization of a two-time correlation function when $t_2-t_1\ge \tau_c$, i.e., $t_1$ and $t_2$ differ by at least the memory time $\tau_c$. Propagation over $2\tau_c^2$ contains all information needed to reconstruct two-time correlations with arbitrary time arguments.}
	\end{figure}

	The resulting stationary dynamical map $\mathcal{E}^S_{\Delta t}=\mathcal{E}_{\tau_c+\Delta t, \tau_c}$ was used~\cite{timelocal} to extrapolate short-time non-Markovian open quantum system dynamics to long times by $\mathcal{E}_{t,t_0}=\mathcal{E}^S_{t-t_0-\tau_c}\mathcal{E}_{t_0+\tau_c,t_0}$ using $\mathcal{E}^S_{(m\Delta t)}=(\mathcal{E}^S_{\Delta t})^m$ [see Fig.~\ref{fig:fac_two}(a)]. A similar line of reasoning also enables the factorization of multitime correlations.\smallskip
	
	\textit{Factorization rule for correlators -- } We consider the generic $n$-time correlators
	\begin{align}
		\label{eq:basic}
		\langle \mathbf{X}(t_n) A^{(n-1)}(t_{n-1}) \dots A^{(1)}(t_1)
		\rangle_{\mathbf{X}(t_0)},
	\end{align}
	which are to be understood as superoperators on the Liouville space, on which the density matrix is a vector. $\mathbf{X}$ is the vector whose entries are the complete basis of operators $|i\rangle\langle j|$, and the subscript $\mathbf{X}$ indicates that $n$-time correlations are also evaluated for a complete set of initial states $|i\rangle\langle j|$, assuming no correlations between system and environment at time $t_0$.
	
	For example, this notation allows us to express dynamical maps as 
	$\mathcal{E}_{t,t_0}= \langle \mathbf{X}(t) \rangle_{\mathbf{X}(t_0)}$,
	while two-time correlations can be written as
	\begin{equation}
		\langle A^{(2)}(t_2)A^{(1)}(t_1)\rangle 
		=\textrm{Tr} \big\{ A^{(2)}\,
		\langle \mathbf{X}(t_2) A^{(1)}(t_1)
		\rangle_{\mathbf{X}(t_0)} \, \rho(t_0)\big\},
	\end{equation}
	where $\textrm{Tr}$ indicates a scalar product on Liouville space with the 
	same effect as the trace in Hilbert space \cite{gyamfi2020fundamentals}. 
	
	The key result, derived in appendix~\ref{app:deriv}, is that correlators of the form of Eq.~\eqref{eq:basic} with time differences of subsequent operators $t_{j}-t_{j-1}$ exceeding the memory time $\tau_c$ of the environment can be split into smaller correlators, according to the rule
	\begin{align}
		\label{eq:rule}
		&\langle \mathbf{X}(t_n) A^{(n-1)}(t_{n-1}) \dots A^{(1)}(t_1)
		\rangle_{\mathbf{X}(t_0)}
		\nonumber\\&
		=\langle \mathbf{X}(t_n) A^{(n-1)}(t_{n-1}) \dots A^{(j)}(t_j)
		\rangle_{\mathbf{X}(t_j-\tau_c)} 
		\,\mathcal{E}^{\text{int}}_{t_j-t_{j-1}-2\tau_c}\,
		\nonumber\\& \phantom{=} \times
		\langle \mathbf{X}(t_{j-1}+\tau_c) A^{(j-1)}(t_{j-1}) \dots A^{(1)}(t_1)
		\rangle_{\mathbf{X}(t_0)},
	\end{align} 
	where we define intermediate dynamical maps by
	\begin{align}
		\label{eq:Eint}
		\mathcal{E}^{\text{int}}_{\Delta\tau-2\tau_c}=&
		\mathcal{E}_{\tau_c+t_0,t_0}^{-1}\mathcal{E}^S_{\Delta\tau-\tau_c}.
	\end{align}
	This provides a regression-theorem-like reduction for finite-memory open quantum systems, without invoking any approximation.
	The rule in Eq.~\eqref{eq:rule} is visualized in Fig.~\ref{fig:fac_two}(b) for a two-time correlation function and can be interpreted as follows. Fast transient environment excitations shaken up by the application of operator $A^{(j-1)}$ at time $t_{j-1}$ beyond those captured in stationary dynamical map $\mathcal{E}^{S}$ have died out at time $t_{j-1}+\tau_c$. Thus, $\mathcal{E}^{S}$ correctly describes the evolution from $t_{j-1}+\tau_c$ to $t_j$.
	To prepare a situation where the operator $A^{(j)}$ is applied to the system with transient excitations decayed, we start the propagation of the second correlator with an uncorrelated system and environment at time $t_j-\tau_c$. To undo this additional propagation over time $\tau_c$, we remove it by applying its inverse from the left onto the stationary dynamical map $\mathcal{E}^{S}$ of the prior propagation. 
	This yields a factorization of an $n$-time correlator into an $(n-j+1)$-time correlator and a $j$-time correlator, where the propagation over the time between both correlators is given by the intermediate map $\mathcal{E}^{\text{int}}$. 
	\bigskip
	
	\textit{Consequences -- } 
	From the factorization rule in Eq.~\eqref{eq:rule}, it follows that
	$n$-time correlation functions can be broken down into products involving lower-order correlators, unless all time differences $t_{j}-t_{j-1}$ (including between first operator time $t_1$ and initial time $t_0$) are smaller than the memory time $\tau_c$. 
	Calculating such an unfactorizable correlator requires a propagation over at most $n\tau_c$. When all unfactorizable $n$-time correlators are arranged in a hypercube with axes corresponding to time differences $t_j-t_{j-1}$, they fit into a temporal volume of $\tau_c^n$. 
	
	This suggests that $n$-time correlation functions for processes whose overall time scale $t^\text{tot}\gg \tau_c$ far exceeds the memory time of the environment can be calculated extremely efficiently. To this end, one needs to employ accurate yet numerically costly algorithms for non-Markovian open quantum system simulations only for obtaining the unfactorizable $m$-time correlators for $m\le n$. Then, $n$-time correlation functions with arbitrary time arguments can be obtained using the factorization rule.
	As demonstrated below, an eigenvector decomposition of the stationary dynamical map is particularly useful because it facilitates semianalytical solutions and gives direct access to stationary states. Moreover, we show how the factorization rules can be used for systems driven by short time-dependent pulses. \smallskip
	\begin{figure}
		\begin{center}
			\includegraphics[width=0.99\linewidth]{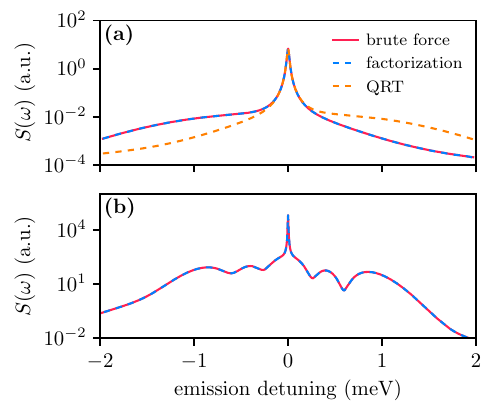}
		\end{center}
		\caption{\label{fig:spectra}(a) Emission spectrum of a weakly cw-driven QD with non-Markovian coupling to phonons. Standard QRT predicts the phonon sideband on the wrong side of the zero-phonon line. Using factorization the complete information is contained in a temporal volume of $\tau_c^2$. (b) Exemplary emission spectrum for resonant pulsed excitation. 
		}
	\end{figure}
	
	\textit{Example: Emission spectra of QDs -- }
	An ideal test case for the factorization rule is the emission spectrum of a semiconductor QD. Due to the strong coupling to phonons the conventional QRT is known to fail as it predicts phonon sidebands on the wrong side of the zero-phonon line~\cite{nonMarkovian_McCutcheon,QRT_Cosacchi}. We first consider a QD under weak cw driving coupled to a super-Ohmic phonon bath with established parameters~\cite{krummheuer2005pure} (see Appendix \ref{app:params}). 
	The emission spectrum is described by
	\begin{equation}
		\label{eq:spectrum_cw}
		S(\omega)=\text{Re}\lim_{t\to\infty}\int\limits_{0}^\infty d\tau\,
		\langle \sigma^+(\tau+t) \sigma^-(t)\rangle e^{-(i\omega+\Gamma) \tau},
	\end{equation}
	where $\sigma^\pm$ are raising and lowering operators of the two-level system comprised of the ground and exciton state of the QD. The limit $t\to\infty$ indicates that the spectrum is measured from the stationary state of the system. Note that the overall time scale $t^\text{tot}$ is set by the radiative decay time of the QD on the order of nanoseconds, while the phonon dynamics and the memory time is on the picosecond scale. 
	
	The brute-force solution is to simulate the system starting from time $t_0=0$ and solve the non-Markovian dynamics up to time $t_1=t^\text{tot}/2$, where the operator $\sigma^-$ is applied. This long time ensures that the stationary state is reached before $t_1$. Then the system is propagated further up to time $t_2=t^\text{tot}$ while extracting the expectation value of $\sigma^+$. The result is transformed according to Eq.~\eqref{eq:spectrum_cw} to yield the spectrum.\\
	The factorization rule enables shortcuts based on the eigenvector decomposition of the stationary dynamical map $\mathcal{E}^S_\tau = \sum_\nu \mathbf{v}_\nu e^{z_\nu \tau} \tilde{\mathbf{v}}^\dagger_\nu$, where $\mathbf{v}_\nu$ and $\tilde{\mathbf{v}}_\nu$ form a biorthogonal basis of right and left eigenvectors. $z_\nu$ is the eigenvalue of the corresponding stationary Liouvillian. We denote the steady state of the system by $\mathbf{v}_0$, for which $z_{0}=0$.\\
	First, because $t_1-t_0=t^\text{tot}/2>\tau_c$, we can separate the two-time correlator from the propagation of the initial state [as in Fig.~\ref{fig:fac_two}(b)]. But instead of applying the intermediate map for a long time duration, one may take
	the steady state $\mathbf{v}_0$ from the eigenvector decomposition of the stationary map. Just like the intermediate map, this state has to be backpropagated to yield a suitable initial state $\mathbf{v}^\text{init}=\mathcal{E}_{\tau_c+t_0,t_0}^{-1} \mathbf{v}_0 $ for the subsequent two-time correlator. We find
	\begin{align}
		&\lim_{t\to\infty}\langle \sigma^+(\tau+t)\sigma^-(t)\rangle
		\nonumber\\&
		=\begin{cases}
			\text{Tr}\left\{\,\sigma^+ \,\langle \mathbf{X}(\tau) \sigma^-(0)\rangle_{\mathbf{v}^\text{init}(-\tau_c)}\right\}, & \tau<\tau_c, \\
			\text{Tr}\left\{\,\sigma^+ \mathcal{E}^S_{\tau-\tau_c} \,\langle \mathbf{X}(\tau_c) \sigma^-(0) \rangle_{\mathbf{v}^\text{init}(-\tau_c)}\right\}, & \tau\ge\tau_c,
		\end{cases}
	\end{align}
	where we have shifted absolute times so that $t_1\to 0$.
	
	To evaluate the spectrum, we separately consider the contributions from $\tau<\tau_c$ and $\tau\ge\tau_c$ to the integral in Eq.~\eqref{eq:spectrum_cw}, which yields
	\begin{align}
		&S(\omega)= \text{Re}
		\int\limits_{0}^{\tau_c} d\tau\,\text{Tr}\big\{ \sigma^+ \,\langle \mathbf{X}(\tau) \sigma^-(0)\rangle_{\mathbf{v}^\text{init}(-\tau_c)} e^{-(i\omega+\Gamma) \tau}\big\} \nonumber\\
		&\!+\!\sum_\nu \!\text{Tr} \big\{ \sigma^+ \mathbf{v}_\nu\big\}  f_{\tau_c}(z_\nu\!-\!i\omega\!-\!\Gamma)\! 
		\big( \tilde{\mathbf{v}}^\dagger_\nu  \langle \mathbf{X}(\tau_c) \sigma^{-}(0) \rangle_{\mathbf{v}^\text{init}(-\tau_c)} \big),
	\end{align}
	where $f_{\tau_c}(z_\nu-i\omega-\Gamma)=\int_{\tau_c}^\infty d\tau e^{(z_\nu-i\omega-\Gamma) \tau}$
	are analytically solvable. For $\text{Re}(z_\nu)<0$ they yield Lorentzian lines, corresponding to resonance fluorescence~\cite{Mollow}. Purely imaginary eigenvalues $\text{Re}(z_\nu)=0$ contribute as Dirac $\delta$ functions, which describe elastic scattering~\cite{Mollow}. While these lines may experience renormalization and broadening due to the environment, non-Lorentzian phonon sidebands have to stem from non-Markovian correlations within the memory time $\tau<\tau_c$. 
	
	In Fig.~\ref{fig:spectra}(a), simulated emission spectra are shown for a QD coupled to phonons with parameters given in Appendix~\ref{app:params}. These spectra are obtained using a numerically exact process tensor framework~\cite{ACE,DnC,ACE_code} using brute-force simulation over a long time span $t^\text{tot}$ (red) as well as using the factorization rule (blue dashed), where the two-time correlator is calculated only over time $2\tau_c$. The results match exactly, while the factorization method leads to a speed-up of two to three orders of magnitude depending on $t^{\mathrm{tot}}$ and $\tau_c$. For comparison, the QRT (orange) yields qualitatively different results.\smallskip
	
	\textit{Pulsed excitation: -- }
	Recently, the question of emission spectra under pulsed excitation received
	attention due to the interesting features arising from interfering Mollow
	lines~\cite{boos2024signatures,liu2024dynamic}. This situation deviates from the previous example in that the total Hamiltonian $H_0+H_d(t)$ acquires a time dependent driving term, and the measured signal is proportional to the spectrum integrated over the first time $t$
	\begin{align}
		S(\omega)=\mathrm{Re}\int\limits_0^{\infty}dt\int\limits_0^{\infty} d\tau 
		\langle \sigma^+(\tau+t) \sigma^-(t)\rangle e^{-i\omega \tau}.
	\end{align}
	The time dependence of $H_d(t)$ seems to break the assumption of our decomposition rule. However, this rule can be generalized if $H_d(t)$ is nonzero only over a finite time interval $[t_j, t_j+\Delta t_j]$. Just like the application of an operator $A^{(i)}$ in multitime correlations constitutes an instantaneous intervention, a pulse can be regarded as an intervention 
	\begin{align}
		P^{(j)}=e^{\frac{i}{\hbar}H_0\Delta t_j} 
		\mathcal{T}e^{-\frac{i}{\hbar}\int_{t_j}^{t_j+\Delta t_j}(H_0+H_d(t'))dt'}
		\label{eq:pulse_intervention}
	\end{align}
	with finite duration $\Delta t_j$, which effectively prolongs the time duration for environment excitations to relax from $\tau_c$ to $\tau_c+\Delta t_j$. A straightforward generalization of Eq.~\eqref{eq:rule} lets us then express the contribution to the time-integrated spectrum for $t>\Delta t_0+\tau_c$ and $\tau>\tau_c$ as 
	\begin{widetext}
		\begin{align}
			&\int\limits_{\Delta t_0+\tau_c}^{\infty}\!\!\!dt\int\limits_{\tau_c}^{\infty} d\tau  \langle \sigma^+(\tau+t) \sigma^-(t)\rangle e^{-i\omega \tau}
			\nonumber\\
			&=\int\limits_{0}^{\infty}dt'\int\limits_{\tau_c}^{\infty} d\tau\,\text{Tr}_S\big\{\sigma^+ \mathcal{E}^{S}_{\tau-\tau_c} \langle \mathbf{X}(2\tau_c) \sigma^-(\tau_c)\rangle_\mathbf{X}\mathcal{E}^\text{int}_{t'} \langle \mathbf{X}(\Delta t_0 + \tau_c) P^{(0)}(t_0)\rangle_{\rho(t_0)}\big\}e^{-i\omega\tau}\nonumber\\
			&=\sum_{\nu,\nu'} \text{Tr}_S\big\{\sigma^+ \mathbf{v}_{\nu'}\big\}\big(\tilde{\mathbf{v}}^\dagger_{\nu'}\langle \mathbf{X}(2\tau_c) \sigma^-(\tau_c)\rangle_\mathbf{X}(\mathcal{E}_{\tau_c+t_0,t_0})^{-1} \mathbf{v}_{\nu}\big)\big(\tilde{\mathbf{v}}^\dagger_{\nu} \langle \mathbf{X}(\Delta t_0 + \tau_c) P^{(0)}(t_0)\rangle_{\rho(t_0)}\big)
			f_{0}(z_\nu)f_{\tau_c}(z_{\nu'}-i\omega),
		\end{align}
	\end{widetext}
	where in the second line we relabeled $t'=t-(\Delta t_0+\tau_c)$.
	Again, the time integrals are solved analytically and yield Lorentzians. 
	The largest unfactorizable contribution spans the time intervals $t\in[0, \Delta t_0+\tau_c]$ and $\tau\in[0,\tau_c]$, i.e., a temporal volume of $(\Delta t_0+\tau_c)\tau_c$. This is in contrast to a direct treatment, which requires integration over a temporal volume $\mathcal{O}(T_\infty^2)$. Despite these savings, time-integrated spectra for pulsed excitation are exactly reproduced, as can be seen in Fig.~\ref{fig:spectra}(b). Parameters are given in Appendix~\ref{app:params}.\\
	
\begin{figure}
		\begin{center}
			\includegraphics[width=0.99\linewidth]{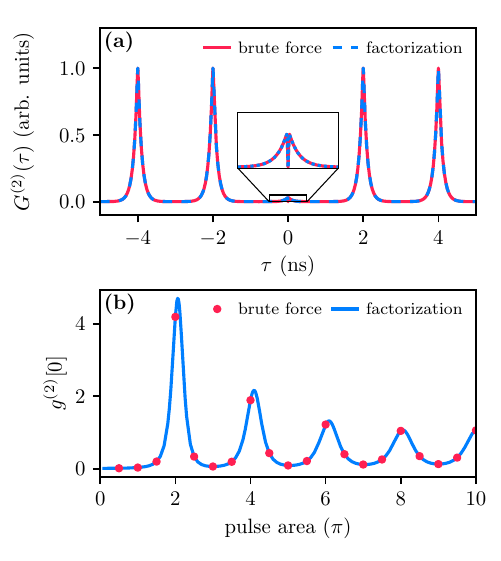}
		\end{center}
		\caption{\label{fig:purity}(a) Time-integrated second-order coherence for a resonant pulse with $\sigma=\SI{5}{ps}$ and pulse area of $\pi$, resulting in $g^{(2)}[0]=0.032$. For other parameters, see appendix. (b) Ratio of photon counts stemming from a single pulse and subsequent pulses at different pulse areas.}
	\end{figure}

	\textit{Purity: -- }
	Important figures of merit for quantum light sources, such as purity, 
	indistinguishability, and concurrence, are based on second-order coherences like $G^{(2)}(t,\tau)=\langle \sigma^+(t)\sigma^+(t+\tau)\sigma^-(t+\tau)\sigma^-(t)\rangle$ \cite{fischer2016dynamical,hanschke2018quantum}. Here, we consider concretely the purity of single photons, quantified by the integrated second-order coherence $g^{(2)}[0]$. For this, a QD is excited using a pulse train with period $T$, and the second-order coherence $G^{(2)}(t,\tau)$ is integrated over time $t$. For the delay time $\tau$ one distinguishes photon pairs originating from a single pulse $\tau\in\left[-\frac{T}{2}, \frac{T}{2}\right]$ and from subsequent pulses, e.g., $\tau\in\left[\frac{T}{2}, \frac{3T}{2}\right]$. In general $T\gg 1/\gamma$ is chosen, where $1/\gamma$ is the exciton lifetime, such that in between pulses the QD relaxes completely to its ground state. Here, we use $1/\gamma=\SI{100}{ps}$ and $T=\SI{2}{ns}$.\\
	As the pulses act only briefly during each excitation period, most of the time the total Hamiltonian is time-independent. We can again assume any pulse as an intervention $P^{(j)}$ like given in Eq.~\eqref{eq:pulse_intervention}. If pulses in each period of the pulse train are identical, all $P^{(j)}$ are the same and repeat after $t+\tau\ge T$. This allows us to re-use dynamical maps from the first interval of the pulse train for any other combination of $t,\tau$. Fig.~\ref{fig:purity} shows $G^{(2)}(\tau)$ for resonant $\pi$-pulse excitation in panel (a), with a perfect match between brute-force calculation and the usage of correlator factorization. Panel (b) shows $g^{(2)}[0]$, where we calculated the results for pulse areas of integer $\pi$ values using the traditional brute-force method. Using the factorization method allows to calculate hundreds of points in the same amount of time, without sacrificing accuracy. Note that the maxima of $g^{(2)}[0]$ are shifted to slightly higher pulse areas than $2n\pi$ due to finite excited state lifetimes and resulting re-excitation.
	
	\textit{Conclusion --}
	We presented a factorization rule for multitime correlation functions, which is valid even in the presence of strong non-Markovian effects. This technique allows to calculate exact $n$-time correlation functions while only explicitly calculating correlations within a small temporal volume of $\tau_c^n$. As a result, numerical effort can be greatly reduced, in a similar way to how the QRT is conventionally applied to Markovian problems. 
	This factorization rule is available if the environment memory time $\tau_c$ is finite and if the Hamiltonian is time-independent except for short intervals containing finite pulses. 
	We note that the factorization strategy can also be adapted to systems subject to periodic driving, with corresponding time-dependent but periodic dynamical maps and generic correlators.
	For arbitrary time-dependent driving, we believe that no general exact factorization rule can be identified, because arbitrary changes in the Hamiltonian trigger excitations beyond the number of degrees of freedom of the reduced system, and thus prevent equilibration of dynamical maps.
	Nevertheless, we expect that even in this case, our factorization rule is a promising starting point for suitable approximate estimates for multitime correlations in strongly non-Markovian open quantum systems.
	
	\section*{Acknowledgments}
	The authors acknowledge financial support of the Return Program of the State of North Rhine‑Westphalia.
	\bibliography{references}

@article{budini2008operator,
	author = {Budini, Adri{\'a}n A. },
	doi = {10.1007/s10955-007-9476-9},
	id = {Budini2008},
	isbn = {1572-9613},
	journal = {J. Stat. Phys.},
	number = {1},
	pages = {51--78},
	title = {Operator Correlations and Quantum Regression Theorem in Non-Markovian Lindblad Rate Equations},
	url = {https://doi.org/10.1007/s10955-007-9476-9},
	volume = {131},
	year = {2008},
}

@misc{bundgaardnielsen2026quantum,
      title={Beyond the Quantum Regression Theorem in Variational Polaron Master Equations with Low-Dimensional Baths}, 
      author={Matias Bundgaard-Nielsen and Jake Iles-Smith},
      year={2026},
      eprint={2604.13541},
      archivePrefix={arXiv},
      primaryClass={quant-ph},
      doi={10.48550/arXiv.2604.13541},
      url={https://arxiv.org/abs/2604.13541}, 
}

@article{alonso2005multiple,
  title = {Multiple-Time Correlation Functions for Non-Markovian Interaction: Beyond the Quantum Regression Theorem},
  author = {Alonso, Daniel and de Vega, In\'es},
  journal = {Phys. Rev. Lett.},
  volume = {94},
  issue = {20},
  pages = {200403},
  numpages = {4},
  year = {2005},
  month = {May},
  publisher = {American Physical Society},
  doi = {10.1103/PhysRevLett.94.200403},
  url = {https://link.aps.org/doi/10.1103/PhysRevLett.94.200403}
}

@article{khan2022quantum,
  title = {Quantum regression theorem for multi-time correlators: A detailed analysis in the Heisenberg picture},
  author = {Khan, Sakil and Agarwalla, Bijay Kumar and Jain, Sachin},
  journal = {Phys. Rev. A},
  volume = {106},
  issue = {2},
  pages = {022214},
  numpages = {16},
  year = {2022},
  month = {Aug},
  publisher = {American Physical Society},
  doi = {10.1103/PhysRevA.106.022214},
  url = {https://link.aps.org/doi/10.1103/PhysRevA.106.022214}
}

@article{Eberly,
author = {J. H. Eberly and K. W\'{o}dkiewicz},
journal = {J. Opt. Soc. Am.},
number = {9},
pages = {1252--1261},
publisher = {Optica Publishing Group},
title = {The time-dependent physical spectrum of light$\ast$},
volume = {67},
month = {Sep},
year = {1977},
url = {https://opg.optica.org/abstract.cfm?URI=josa-67-9-1252},
doi = {10.1364/JOSA.67.001252}
}

@article{QRT_Lax1963,
  title = {Formal Theory of Quantum Fluctuations from a Driven State},
  author = {Lax, Melvin},
  journal = {Phys. Rev.},
  volume = {129},
  issue = {5},
  pages = {2342--2348},
  numpages = {0},
  year = {1963},
  month = {Mar},
  publisher = {American Physical Society},
  doi = {10.1103/PhysRev.129.2342},
  url = {https://link.aps.org/doi/10.1103/PhysRev.129.2342}
}

@article{QRT_Guarnieri,
  title = {Quantum regression theorem and non-Markovianity of quantum dynamics},
  author = {Guarnieri, Giacomo and Smirne, Andrea and Vacchini, Bassano},
  journal = {Phys. Rev. A},
  volume = {90},
  issue = {2},
  pages = {022110},
  numpages = {11},
  year = {2014},
  month = {Aug},
  publisher = {American Physical Society},
  doi = {10.1103/PhysRevA.90.022110},
  url = {https://link.aps.org/doi/10.1103/PhysRevA.90.022110}
}

@article{QRT_Swain,
doi = {10.1088/0305-4470/14/10/013},
url = {https://doi.org/10.1088/0305-4470/14/10/013},
year = {1981},
month = {oct},
publisher = {},
volume = {14},
number = {10},
pages = {2577},
author = {S Swain},
title = {Master equation derivation of quantum regression theorem},
journal = {J. Phys. A: Math. Gen.}
}

@article{QRT_FordOConnell,
  title = {There is No Quantum Regression Theorem},
  author = {Ford, G. W. and O'Connell, R. F.},
  journal = {Phys. Rev. Lett.},
  volume = {77},
  issue = {5},
  pages = {798--801},
  numpages = {0},
  year = {1996},
  month = {Jul},
  publisher = {American Physical Society},
  doi = {10.1103/PhysRevLett.77.798},
  url = {https://link.aps.org/doi/10.1103/PhysRevLett.77.798}
}

@article{QRT_Cosacchi,
  title = {Accuracy of the Quantum Regression Theorem for Photon Emission from a Quantum Dot},
  author = {Cosacchi, M. and Seidelmann, T. and Cygorek, M. and Vagov, A. and Reiter, D. E. and Axt, V. M.},
  journal = {Phys. Rev. Lett.},
  volume = {127},
  issue = {10},
  pages = {100402},
  numpages = {8},
  year = {2021},
  month = {Aug},
  publisher = {American Physical Society},
  doi = {10.1103/PhysRevLett.127.100402},
  url = {https://link.aps.org/doi/10.1103/PhysRevLett.127.100402}
}

@article{BLP,
  title = {Measure for the Degree of Non-Markovian Behavior of Quantum Processes in Open Systems},
  author = {Breuer, Heinz-Peter and Laine, Elsi-Mari and Piilo, Jyrki},
  journal = {Phys. Rev. Lett.},
  volume = {103},
  issue = {21},
  pages = {210401},
  numpages = {4},
  year = {2009},
  month = {Nov},
  publisher = {American Physical Society},
  doi = {10.1103/PhysRevLett.103.210401},
  url = {https://link.aps.org/doi/10.1103/PhysRevLett.103.210401}
}

@article{RHP,
  title = {Entanglement and Non-Markovianity of Quantum Evolutions},
  author = {Rivas, \'Angel and Huelga, Susana F. and Plenio, Martin B.},
  journal = {Phys. Rev. Lett.},
  volume = {105},
  issue = {5},
  pages = {050403},
  numpages = {4},
  year = {2010},
  month = {Jul},
  publisher = {American Physical Society},
  doi = {10.1103/PhysRevLett.105.050403},
  url = {https://link.aps.org/doi/10.1103/PhysRevLett.105.050403}
}

@misc{timelocal,
      title={Time-nonlocal versus time-local long-time extrapolation of non-Markovian quantum dynamics}, 
      author={Moritz Cygorek and Erik M. Gauger},
      year={2025},
      eprint={2505.21017},
      archivePrefix={arXiv},
      primaryClass={quant-ph},
      url={https://arxiv.org/abs/2505.21017}, 
}

@article{timelocal_Clark,
    author = {Strachan, David J. and Purkayastha, Archak and Clark, Stephen R.},
    title = {Extracting dynamical maps of non-Markovian open quantum systems},
    journal = {J. Chem. Phys.},
    volume = {161},
    number = {15},
    pages = {154105},
    year = {2024},
    month = {10},
    issn = {0021-9606},
    doi = {10.1063/5.0228428},
    url = {https://doi.org/10.1063/5.0228428}
}

@article{nonMarkovian_McCutcheon,
  title = {Optical signatures of non-Markovian behavior in open quantum systems},
  author = {McCutcheon, Dara P. S.},
  journal = {Phys. Rev. A},
  volume = {93},
  issue = {2},
  pages = {022119},
  numpages = {7},
  year = {2016},
  month = {Feb},
  publisher = {American Physical Society},
  doi = {10.1103/PhysRevA.93.022119},
  url = {https://link.aps.org/doi/10.1103/PhysRevA.93.022119}
}

@article{krummheuer2005pure,
	title     = {{Pure dephasing and phonon dynamics in GaAs- and GaN-based quantum dot structures:
	Interplay between material parameters and geometry}},
	author    = {Krummheuer, B. and Axt, V. M. and Kuhn, T. and D'Amico, I. and Rossi, F.},
	journal   = {Phys. Rev. B},
	volume    = {71},
	pages     = {235329},
	year      = {2005},
	doi		  = {10.1103/PhysRevB.71.235329}
}

@article{PImultitime,
  title = {Path-integral approach for nonequilibrium multitime correlation functions of open quantum systems coupled to Markovian and non-Markovian environments},
  author = {Cosacchi, M. and Cygorek, M. and Ungar, F. and Barth, A. M. and Vagov, A. and Axt, V. M.},
  journal = {Phys. Rev. B},
  volume = {98},
  issue = {12},
  pages = {125302},
  numpages = {10},
  year = {2018},
  month = {Sep},
  publisher = {American Physical Society},
  doi = {10.1103/PhysRevB.98.125302},
  url = {https://link.aps.org/doi/10.1103/PhysRevB.98.125302}
}

@article{Kiraz,
  title = {Quantum-dot single-photon sources: Prospects for applications in linear optics quantum-information processing},
  author = {Kiraz, A. and Atat\"ure, M. and Imamo\ifmmode \breve{g}\else \u{g}\fi{}lu, A.},
  journal = {Phys. Rev. A},
  volume = {69},
  issue = {3},
  pages = {032305},
  numpages = {10},
  year = {2004},
  month = {Mar},
  publisher = {American Physical Society},
  doi = {10.1103/PhysRevA.69.032305},
  url = {https://link.aps.org/doi/10.1103/PhysRevA.69.032305}
}

@article{twophoton,
  title = {Quantum correlations of spontaneous two-photon emission from a quantum dot},
  volume = {643},
  ISSN = {1476-4687},
  url = {http://dx.doi.org/10.1038/s41586-025-09267-6},
  DOI = {10.1038/s41586-025-09267-6},
  number = {8074},
  journal = {Nature},
  publisher = {Springer Science and Business Media LLC},
  author = {Liu,  Shunfa and Wang,  Yangpeng and Saleem,  Yasser and Li,  Xueshi and Liu,  Hanqing and Yang,  Cheng-Ao and Yang,  Jiawei and Ni,  Haiqiao and Niu,  Zhichuan and Meng,  Yun and Hu,  Xiaolong and Yu,  Ying and Wang,  Xuehua and Cygorek,  Moritz and Liu,  Jin},
  year = {2025},
  month = jul,
  pages = {1234–1239}
}

@article{IlesSmith2026,
  title = {Markovian Approach to $N$-Photon Correlations beyond the Quantum Regression Theorem},
  author = {Salamon, Mateusz and Dudgeon, Oliver and Nazir, Ahsan and Iles-Smith, Jake},
  journal = {Phys. Rev. Lett.},
  volume = {136},
  issue = {8},
  pages = {080401},
  numpages = {6},
  year = {2026},
  month = {Feb},
  publisher = {American Physical Society},
  doi = {10.1103/pgk3-b57l},
  url = {https://link.aps.org/doi/10.1103/pgk3-b57l}
}

@article{Mollow,
  title = {Power Spectrum of Light Scattered by Two-Level Systems},
  author = {Mollow, B. R.},
  journal = {Phys. Rev.},
  volume = {188},
  issue = {5},
  pages = {1969--1975},
  numpages = {0},
  year = {1969},
  month = {Dec},
  publisher = {American Physical Society},
  doi = {10.1103/PhysRev.188.1969},
  url = {https://link.aps.org/doi/10.1103/PhysRev.188.1969}
}

@article{ACE,
author={Cygorek, Moritz
and Cosacchi, Michael
and Vagov, Alexei
and Axt, Vollrath Martin
and Lovett, Brendon W.
and Keeling, Jonathan
and Gauger, Erik M.},
title={Simulation of open quantum systems by automated compression of arbitrary environments},
journal={Nat. Phys.},
year={2022},
month={Jun},
day={01},
volume={18},
number={6},
pages={662-668},
issn={1745-2481},
doi={10.1038/s41567-022-01544-9},
url={https://doi.org/10.1038/s41567-022-01544-9}
}

@article{ACE_code,
    author = {Cygorek, Moritz and Gauger, Erik M.},
    title = {ACE: A general-purpose non-Markovian open quantum systems simulation toolkit based on process tensors},
    journal = {J. Chem. Phys.},
    volume = {161},
    number = {7},
    pages = {074111},
    year = {2024},
    month = {08},
    issn = {0021-9606},
    doi = {10.1063/5.0221182},
    url = {https://doi.org/10.1063/5.0221182}
}

@article{DnC,
  title = {Sublinear Scaling in Non-Markovian Open Quantum Systems Simulations},
  author = {Cygorek, Moritz and Keeling, Jonathan and Lovett, Brendon W. and Gauger, Erik M.},
  journal = {Phys. Rev. X},
  volume = {14},
  issue = {1},
  pages = {011010},
  numpages = {25},
  year = {2024},
  month = {Feb},
  publisher = {American Physical Society},
  doi = {10.1103/PhysRevX.14.011010},
  url = {https://link.aps.org/doi/10.1103/PhysRevX.14.011010}
}

@article{heindel2023quantumdots,
author = {Tobias Heindel and Je-Hyung Kim and Niels Gregersen and Armando Rastelli and Stephan Reitzenstein},
journal = {Adv. Opt. Photon.},
number = {3},
pages = {613--738},
publisher = {Optica Publishing Group},
title = {Quantum dots for photonic quantum information technology},
volume = {15},
month = {Sep},
year = {2023},
url = {https://opg.optica.org/aop/abstract.cfm?URI=aop-15-3-613},
doi = {10.1364/AOP.490091},
}

@article{gyamfi2020fundamentals,
doi = {10.1088/1361-6404/ab9fdd},
url = {https://doi.org/10.1088/1361-6404/ab9fdd},
year = {2020},
month = {oct},
publisher = {IOP Publishing},
volume = {41},
number = {6},
pages = {063002},
author = {Gyamfi, Jerryman A},
title = {Fundamentals of quantum mechanics in Liouville space},
journal = {Eur. J. Phys.},
}

@article{boos2024signatures,
	title = {Signatures of Dynamically Dressed States},
	author = {Boos, Katarina and Kim, Sang Kyu and Bracht, Thomas and Sbresny, Friedrich and Kaspari, Jan M. and Cygorek, Moritz and Riedl, Hubert and Bopp, Frederik W. and Rauhaus, William and Calcagno, Carolin and Finley, Jonathan J. and Reiter, Doris E. and M\"uller, Kai},
	journal = {Phys. Rev. Lett.},
	volume = {132},
	number = {5},
	pages = {053602},
	numpages = {7},
	year = {2024},
	publisher = {American Physical Society},
	doi = {10.1103/PhysRevLett.132.053602},
}

@article{liu2024dynamic,
	author = {Liu, Shunfa and Gustin, Chris and Liu, Hanqing and Li, Xueshi and Yu, Ying and Ni, Haiqiao and Niu, Zhichuan and Hughes, Stephen and Wang, Xuehua and Liu, Jin},
	date = {2024/04/01},
	doi = {10.1038/s41566-023-01359-x},
	id = {Liu2024},
	isbn = {1749-4893},
	journal = {Nat. Photon.},
	number = {4},
	pages = {318--324},
	title = {Dynamic resonance fluorescence in solid-state cavity quantum electrodynamics},
	url = {https://doi.org/10.1038/s41566-023-01359-x},
	volume = {18},
	year = {2024}}

@article{fischer2016dynamical,
	title={Dynamical modeling of pulsed two-photon interference},
	author={Fischer, Kevin A and M{\"u}ller, Kai and Lagoudakis, Konstantinos G and Vu{\v{c}}kovi{\'c}, Jelena},
	journal={New J. Phys.},
	volume={18},
	number={11},
	pages={113053},
	year={2016},
	doi={10.1088/1367-2630/18/11/113053},
	publisher={IOP Publishing}
}

@article{reiter2019distinctive,
author = {D. E. Reiter and T. Kuhn and V. M. Axt},
title = {Distinctive characteristics of carrier-phonon interactions in optically driven semiconductor quantum dots},
journal = {Adv. Phys.: X},
volume = {4},
number = {1},
pages = {1655478},
year  = {2019},
publisher = {Taylor & Francis},
doi = {10.1080/23746149.2019.1655478},
}

@article{ilessmith2017limits,
  title = {Limits to coherent scattering and photon coalescence from solid-state quantum emitters},
  author = {Iles-Smith, Jake and McCutcheon, Dara P. S. and M\o{}rk, Jesper and Nazir, Ahsan},
  journal = {Phys. Rev. B},
  volume = {95},
  issue = {20},
  pages = {201305(R)},
  numpages = {6},
  year = {2017},
  month = {May},
  publisher = {American Physical Society},
  doi = {10.1103/PhysRevB.95.201305},
  url = {https://link.aps.org/doi/10.1103/PhysRevB.95.201305}
}

@article{ilessmith2017phonon,
	author = {Iles-Smith, Jake and McCutcheon, Dara P. S. and Nazir, Ahsan and M{\o}rk, Jesper},
	date = {2017/08/01},
	doi = {10.1038/nphoton.2017.101},
	id = {Iles-Smith2017},
	isbn = {1749-4893},
	journal = {Nature Photon.},
	number = {8},
	pages = {521--526},
	title = {Phonon scattering inhibits simultaneous near-unity efficiency and indistinguishability in semiconductor single-photon sources},
	url = {https://doi.org/10.1038/nphoton.2017.101},
	volume = {11},
	year = {2017},
}

@article{stauber2000electronphonon,
  title = {Electron-phonon interaction in quantum dots: A solvable model},
  author = {Stauber, T. and Zimmermann, R. and Castella, H.},
  journal = {Phys. Rev. B},
  volume = {62},
  issue = {11},
  pages = {7336--7343},
  numpages = {0},
  year = {2000},
  month = {Sep},
  publisher = {American Physical Society},
  doi = {10.1103/PhysRevB.62.7336},
  url = {https://link.aps.org/doi/10.1103/PhysRevB.62.7336}
}

@article{chruscinski2022dynamical,
    title = {Dynamical maps beyond Markovian regime},
    journal = {Phys. Rep.},
    volume = {992},
    pages = {1-85},
    year = {2022},
    note = {Dynamical maps beyond Markovian regime},
    issn = {0370-1573},
    doi = {10.1016/j.physrep.2022.09.003},
    author = {Dariusz Chruściński},
}

@article{glauber1963quantum,
  title = {The Quantum Theory of Optical Coherence},
  author = {Glauber, Roy J.},
  journal = {Phys. Rev.},
  volume = {130},
  issue = {6},
  pages = {2529--2539},
  numpages = {0},
  year = {1963},
  month = {Jun},
  publisher = {American Physical Society},
  doi = {10.1103/PhysRev.130.2529},
  url = {https://link.aps.org/doi/10.1103/PhysRev.130.2529}
}

@article{mcalister1997ultra,
  title = {Ultrafast photon-number correlations from dual-pulse, phase-averaged homodyne detection},
  author = {McAlister, D. F. and Raymer, M. G.},
  journal = {Phys. Rev. A},
  volume = {55},
  issue = {3},
  pages = {R1609(R)--R1612(R)},
  numpages = {0},
  year = {1997},
  month = {Mar},
  publisher = {American Physical Society},
  doi = {10.1103/PhysRevA.55.R1609},
  url = {https://link.aps.org/doi/10.1103/PhysRevA.55.R1609}
}

@article{reiter2014roleofphonons,
doi = {10.1088/0953-8984/26/42/423203},
url = {https://doi.org/10.1088/0953-8984/26/42/423203},
year = {2014},
month = {oct},
publisher = {IOP Publishing},
volume = {26},
number = {42},
pages = {423203},
author = {Reiter, D E and Kuhn, T and Glässl, M and Axt, V M},
title = {The role of phonons for exciton and biexciton generation in an optically driven quantum dot},
journal = {J. Phys.: Condens. Matter},
}

@article{hanschke2018quantum,
	author = {Hanschke, Lukas and Fischer, Kevin A. and Appel, Stefan and Lukin, Daniil and Wierzbowski, Jakob and Sun, Shuo and Trivedi, Rahul and Vu{\v c}kovi{\'c}, Jelena and Finley, Jonathan J. and M{\"u}ller, Kai},
	doi = {10.1038/s41534-018-0092-0},
	id = {Hanschke2018},
	isbn = {2056-6387},
	journal = {npj Quantum Inf.},
	number = {1},
	pages = {43},
	title = {Quantum dot single-photon sources with ultra-low multi-photon probability},
	url = {https://doi.org/10.1038/s41534-018-0092-0},
	volume = {4},
	year = {2018},
}

@article{mcalister1997ultrafast,
  title = {Ultrafast photon-number correlations from dual-pulse, phase-averaged homodyne detection},
  author = {McAlister, D. F. and Raymer, M. G.},
  journal = {Phys. Rev. A},
  volume = {55},
  issue = {3},
  pages = {R1609(R)--R1612(R)},
  numpages = {0},
  year = {1997},
  month = {Mar},
  publisher = {American Physical Society},
  doi = {10.1103/PhysRevA.55.R1609},
}

@article{huber2018semiconductor,
  title={Semiconductor quantum dots as an ideal source of polarization-entangled photon pairs on-demand: a review},
  author={Huber, Daniel and Reindl, Marcus and Aberl, Johannes and Rastelli, Armando and Trotta, Rinaldo},
  journal={J. Opt.},
  volume={20},
  number={7},
  pages={073002},
  year={2018},
  doi={10.1088/2040-8986/aac4c4},
  publisher={IOP Publishing}
}

@article{bracht2024theory,
  title = {Theory of time-bin-entangled photons from quantum emitters},
  author = {Bracht, Thomas K. and Kappe, Florian and Cygorek, Moritz and Seidelmann, Tim and Karli, Yusuf and Remesh, Vikas and Weihs, Gregor and Axt, Vollrath Martin and Reiter, Doris E.},
  journal = {Phys. Rev. A},
  volume = {110},
  issue = {6},
  pages = {063709},
  numpages = {11},
  year = {2024},
  month = {Dec},
  publisher = {American Physical Society},
  doi = {10.1103/PhysRevA.110.063709},
  url = {https://link.aps.org/doi/10.1103/PhysRevA.110.063709}
}

@article{delvalle2012theory,
  title = {Theory of Frequency-Filtered and Time-Resolved $N$-Photon Correlations},
  author = {del Valle, E. and Gonzalez-Tudela, A. and Laussy, F. P. and Tejedor, C. and Hartmann, M. J.},
  journal = {Phys. Rev. Lett.},
  volume = {109},
  issue = {18},
  pages = {183601},
  numpages = {5},
  year = {2012},
  month = {Oct},
  publisher = {American Physical Society},
  doi = {10.1103/PhysRevLett.109.183601},
  url = {https://link.aps.org/doi/10.1103/PhysRevLett.109.183601}
}

@article{rossi2002theory,
  title = {Theory of ultrafast phenomena in photoexcited semiconductors},
  author = {Rossi, Fausto and Kuhn, Tilmann},
  journal = {Rev. Mod. Phys.},
  volume = {74},
  issue = {3},
  pages = {895--950},
  numpages = {0},
  year = {2002},
  month = {Aug},
  publisher = {American Physical Society},
  doi = {10.1103/RevModPhys.74.895},
  url = {https://link.aps.org/doi/10.1103/RevModPhys.74.895}
}

@article{lodahl2016interfacing,
  title = {Interfacing single photons and single quantum dots with photonic nanostructures},
  author = {Lodahl, Peter and Mahmoodian, Sahand and Stobbe, S\o{}ren},
  journal = {Rev. Mod. Phys.},
  volume = {87},
  issue = {2},
  pages = {347--400},
  numpages = {54},
  year = {2015},
  month = {May},
  publisher = {American Physical Society},
  doi = {10.1103/RevModPhys.87.347},
  url = {https://link.aps.org/doi/10.1103/RevModPhys.87.347}
}

@article{akopian2006entangled,
  title = {Entangled Photon Pairs from Semiconductor Quantum Dots},
  author = {Akopian, N. and Lindner, N. H. and Poem, E. and Berlatzky, Y. and Avron, J. and Gershoni, D. and Gerardot, B. D. and Petroff, P. M.},
  journal = {Phys. Rev. Lett.},
  volume = {96},
  issue = {13},
  pages = {130501},
  numpages = {4},
  year = {2006},
  month = {Apr},
  publisher = {American Physical Society},
  doi = {10.1103/PhysRevLett.96.130501},
  url = {https://link.aps.org/doi/10.1103/PhysRevLett.96.130501}
}
	\appendix
	\section{Derivation of factorization rule}\label{app:deriv}
	
	The time evolution of the density matrix $\rho^\text{tot}$ on a Markovian embedding of the open quantum system is given by $\dot{\rho}^\text{tot}=\mathcal{L}\rho^\text{tot}$ with Liouvillian $\mathcal{L}$. The auxiliary degrees of freedom of the Markovian embedding may be, e.g., a reaction coordinate or the microscopic environment space itself. Irrespective of their concrete interpretation, we henceforth refer to them as the environment.
	
	We assume that the Liouvillian is time-independent and we decompose it as $\mathcal{L}=\sum_k \mathbf{r}^{(k)} \lambda^{(k)} \mathbf{l}^{(k)\dagger}$, where $\lambda^{(k)}$ is the $k$-th eigenvalue and $\mathbf{r}^{(k)}$ and $\mathbf{l}^{(k)}$ are the corresponding right and left eigenvectors, respectively, which form a biorthogonal basis $\mathbf{l}^{(k)\dagger}\mathbf{r}^{(k')}=\delta_{k,k'}$ on the composite Liouville space. 
	Then, the time evolution is given by
	\begin{align}
		\rho^\text{tot}(t)=e^{\mathcal{L}t}\rho^\text{tot}(t_0) 
		=\sum_k c^{(k)} e^{\lambda^{(k)}t} \mathbf{r}^{(k)} ,
	\end{align}
	where $c^{(k)}=\mathbf{l}^{(k)\dagger} \rho^\text{tot}(t_0)$ is the component of the initial state along the $k$-th right eigenvector.
	The initial state $\rho^\text{tot}(t_0)=\rho(t_0)\otimes \rho^E(t_0)$ is assumed to be an uncorrelated product of initial reduced system and environment density matrices, $\rho(t_0)$ and $\rho^E(t_0)$, respectively. 
	
	While there are $D^2D_E^2$ Liouvillian eigenvectors, where $D$ and $D_E$ are the respective Hilbert space dimensions of system and environment, only a much smaller number will contribute to the reduced system dynamics~\cite{timelocal}:
	First, the initial state may overlap only with a small subset of right eigenvectors $\mathbf{r}^{(k)}$, and so many coefficients $c^{(k)}\approx 0$ are negligible from the start. Moreover, during the dynamics, components corresponding to eigenvalues $\lambda^{(k)}$ with large negative real parts quickly decay while other contributions cancel due to destructive interference. 
	
	This leads us to following formal definition: We define the memory time $\tau_c$ as the time after which only at most $D^2$ Liouvillian eigenvectors contribute significantly, irrespective of the system initial state ${\rho}(0)$.
	
	For the reduced system density matrix, it follows that
	\begin{align}
		{\rho}(t) =\textrm{Tr}_E\{\rho^\text{tot}(t)\}
		=\sum_k e^{\lambda^{(k)}t} \bar{\mathbf{r}}^{(k)}  c^{(k)},
	\end{align}
	where $\bar{\mathbf{r}}^{(k)}=\textrm{Tr}_E\{\mathbf{r}^{(k)}\}$ are the eigenvectors of the composite Liouvillian reduced to the system subspace. If it is known that a given state has a composite system density matrix $\rho^\text{tot}(t')$ for $t'>\tau_c$ with contributions only along $D^2$ vectors $\mathbf{r}^{(k)}$, it can be formally reconstructed by its reduced system density matrix ${\rho}(t')$ using a dual basis to the reduced Liouville eigenvectors defined by $\bar{\mathbf{l}}^{(k)\dagger} \bar{\mathbf{r}}^{(k')}=\delta_{k,k'}$ via
	\begin{align}
		\rho^\text{tot}(t') =&\sum_k \mathbf{r}^{(k)} c^{(k)}(t')=
		\sum_k \mathbf{r}^{(k)} \big( \bar{\mathbf{l}}^{(k)\dagger}  {\rho}(t')\big).
	\end{align}
	The subsequent time evolution is determined by the Liouvillian eigenvalues. Tracing out the environment yields 
	\begin{align}
		{\rho}(t) =&\bigg[\sum_k \bar{\mathbf{r}}^{(k)} e^{\lambda^{(k)}(t-t')} \bar{\mathbf{l}}^{(k)\dagger}\bigg] {\rho}(t'),
	\end{align}
	where the term in brackets is the dynamical map $\mathcal{E}_{t,t'}$, which becomes exact for $t'>\tau_c$.
	Note that the dynamical map depends only on time differences, and so by introducing a fixed time step $\Delta t$ we arrive at the stationary time-local dynamical map $\mathcal{E}^S_{\Delta t}=\sum_k\bar{\mathbf{r}}^{(k)} e^{\lambda^{(k)}\Delta t} \bar{\mathbf{l}}^{(k)\dagger}$.
	
	To decouple multitime correlation functions, it is useful to formally express the reconstruction of the total density matrix after a time $\tau_c$ has elapsed by
	\begin{align}
		\label{eq:deriv_reconstr}
		e^{\mathcal{L}\tau_c} \dots = & \sum_k \mathbf{r}^{(k)}  \bar{\mathbf{l}}^{(k)\dagger} \text{Tr}_E\big\{e^{\mathcal{L}\tau_c}\dots \big\}
	\end{align}
	Specifically, applying it to the initial environment state gives
	\begin{align}
		e^{\mathcal{L}\tau_c} \big[ \mathbb{1}\otimes \rho^{E}(t_0) \big] =&
		\sum_k \mathbf{r}^{(k)}  \bar{\mathbf{l}}^{(k)\dagger}
		\text{Tr}_E\Big\{e^{\mathcal{L}\tau_c}\big[ \mathbb{1}\otimes \rho^{E}(t_0) \big]\Big\}
		\nonumber\\=&
		\sum_k \mathbf{r}^{(k)}  \bar{\mathbf{l}}^{(k)\dagger} 
		\langle \mathbf{X}(t_0+\tau_c)\rangle_{\mathbf{X}(t_0)}
		\nonumber\\=&
		\sum_k \mathbf{r}^{(k)}  \bar{\mathbf{l}}^{(k)\dagger}
		\mathcal{E}_{t_0+\tau_c, t_0},
	\end{align}
	Multiplying this from the right with the inverse $\mathcal{E}_{t_0+\tau_c, t_0}^{-1}$  allows one to eliminate the explicit reference to Liouville eigenvectors in Eq.~\eqref{eq:deriv_reconstr}
	\begin{align}
		e^{\mathcal{L}\tau_c} \dots = & 
		e^{\mathcal{L}\tau_c} \big[ \mathbb{1}\otimes \rho^{E}(t_0) \big]
		\mathcal{E}_{t_0+\tau_c, t_0}^{-1}
		\text{Tr}_E\big\{e^{\mathcal{L}\tau_c}\dots \big\}.
	\end{align}
	Finally, because for any $\Delta t>0$
	\begin{align}
		&\text{Tr}_E\big[ e^{\mathcal{L}\tau_c} e^{\mathcal{L}\Delta t}  \dots \big] =
		\text{Tr}_E\big[ e^{\mathcal{L}\Delta t}  e^{\mathcal{L}\tau_c} \dots \big]
		\nonumber\\&=
		\text{Tr}_E\big[ e^{\mathcal{L} \Delta t} \sum_k \mathbf{r}^{(k)}\bar{\mathbf{l}}^{(k)\dagger}\big] \text{Tr}_E\big[ e^{\mathcal{L}\tau_c} \dots \big]
		\nonumber\\&=
		\mathcal{E}^S_{\Delta t} \text{Tr}_E\big[ e^{\mathcal{L}\tau_c} \dots \big]
	\end{align}
	we find  for propagation times $t'>\tau$.
	\begin{align}
		\label{eq:app_rule}
		e^{\mathcal{L} t'} \dots = & 
		e^{\mathcal{L}\tau_c} \big[ \mathbb{1}\otimes \rho^{E}(t_0) \big]
		\mathcal{E}_{t_0+\tau_c, t_0}^{-1} \mathcal{E}^S_{t'-\tau_c}
		\text{Tr}_E\big\{e^{\mathcal{L}\tau_c}\dots \big\}.
	\end{align}
	Replacing $e^{\mathcal{L} (t_j-t_{j-1})}$ in a generic $n$-time correlator by Eq.~\eqref{eq:app_rule} yields the factorization rule in the main text. 
	\section{Simulation Parameters}\label{app:params}
	We use standard phonon parameters given in Ref.~\cite{krummheuer2005pure}, and assume spherical QDs with an electron confinement length of \mbox{$a_e=\SI{5}{nm}$} at a temperature of $\SI{4}{K}$. For the memory time of phonons in this scenario, we take $\tau_c=\SI{15}{ps}$, at which point the Frobenius norm of time-local dynamical maps of adjacent time steps differ by less than $10^{-14}$, close to the smallest numerically resolvable difference. \\
	For the spectrum of continuous-wave excitation in Fig.~\ref{fig:spectra}, we use a Rabi frequency of $\hbar\Omega_0 = \SI{0.001}{meV}$, an excited state lifetime of $1/\gamma = \SI{100}{ps}$, and a detector linewidth of $\hbar\Gamma=\SI{0.01}{meV}$. The spectrum of pulsed excitation in panel (b) of the same plot uses a Gaussian shape of the electric field with $\sigma=\SI{3}{ps}$ and a pulse area of $6\pi$.\\
	For the second-order coherences in Fig.~\ref{fig:purity}, we also use Gaussian shaped pulses with $\sigma=\SI{5}{ps}$ and repetition of pulses each $T=\SI{2}{ns}$.
\end{document}